\documentclass[aps,twocolumn,superscriptaddress]{revtex4-1}
\usepackage[colorlinks=true,bookmarks=true,citecolor=magenta,urlcolor=magenta,linkcolor=magenta,breaklinks]{hyperref} % command used
\usepackage{breakurl}
\usepackage{sidecap}
\usepackage{amssymb}
\usepackage{hhline}
\usepackage{urwchancal}
\usepackage{xcolor}
\sidecaptionvpos{figure}{t}
\usepackage{amsmath}
\usepackage{graphicx}
\usepackage{esint}
\usepackage{epstopdf}%This line makes .eps figures into .pdf - please comment out if not required.
\usepackage{rotating}
\graphicspath{{pict/}{./}}
\usepackage{bm}%
\usepackage{txfonts}
\usepackage[classicReIm]{kpfonts}
\begin{document}

\preprint{APS/123-QED}

\title{Scattering and absorption invariance of nonmagnetic particles under duality transformations}

\author{Qingdong Yang}
\affiliation{School of Optical and Electronic Information, Huazhong University of Science and Technology, Wuhan, Hubei 430074, P. R. China}
\author{Weijin Chen}
\affiliation{School of Optical and Electronic Information, Huazhong University of Science and Technology, Wuhan, Hubei 430074, P. R. China}
\author{Yuntian Chen}
\email{yuntian@hust.edu.cn}
\affiliation{School of Optical and Electronic Information, Huazhong University of Science and Technology, Wuhan, Hubei 430074, P. R. China}
\affiliation{Wuhan National Laboratory for Optoelectronics, Huazhong University of Science and Technology, Wuhan, Hubei 430074, P. R. China}
\author{Wei Liu}
\email{wei.liu.pku@gmail.com}
\affiliation{College for Advanced Interdisciplinary Studies, National University of Defense
Technology, Changsha, Hunan 410073, P. R. China}

\begin{abstract}
We revisit the total scatterings (in terms of extinction, scattering and absorption cross sections) by arbitrary clusters of nonmagnetic particles that support optically-induced magnetic responses. Our reexamination is conducted from the perspective of the electromagnetic duality symmetry, and it is revealed that all total scattering properties are invariant under duality transformations. This secures that for self-dual particle clusters, the total scattering properties are polarization independent for any fixed incident direction; while for non-self-dual particle clusters, two scattering configurations that are connected to each other through a duality transformation would exhibit identical scattering properties. This electromagnetic duality induced invariance is irrelevant to specific particle distributions or wave incident directions, which is illustrated for both random and periodic clusters.
\end{abstract}

\maketitle

%\tableofcontents

\section{Introduction}

Similar to other widely explored symmetries of both discrete and continuous natures~\cite{FEYNMAN_2011__Feynmanb}, the electromagnetic duality symmetry between electric and magnetic fields plays an essential role across different disciplines of physics, especially for fundamental studies in \textit{e.g.} singular optics, gauge field theory and string theory~\cite{MISNER_1957_AnnalsofPhysics_Classical,DESER_1976_Phys.Rev.D_Duality,SEN_1993_NuclearPhysicsB_Electricmagnetic,HUANG_2007__Fundamental,BERRY_2017__HalfCentury,HSIEH_2019_Phys.Rev.Lett._Anomaly}. Though it is shown clearly that electromagnetic duality can render great flexibilities for manipulations of light-matter interactions~\cite{FERNANDEZ-CORBATON_2013_Phys.Rev.Lett._Electromagnetica,FERNANDEZ-CORBATON_2013_Phys.Rev.B_Role},  such a symmetry still resists to be widely employed for practical photonic applications, simply because such an employment requires not only electric but also equally strong magnetic responses. Nevertheless, intrinsic magnetism is naturally rare and weak, especially in higher frequency regimes, which makes the barrier of the wide exploitation of the duality symmetry in optics and photonics almost insurmountable~\cite{JACKSON_1998__Classical}.

A feasible solution to the aforementioned problem comes from the central concept in the fields of metamaterials and metasurfaces, that is optically-induced magnetism originating from displacement currents in nonmagnetic structures~\cite{Pendry1999_ITMT,Cai2010_book,CHEN_Rep.Prog.Phys._review_2016}. This core concept has recently penetrated various branches of optics and photonics, and spawned vibrant fields including all-dielectric meta-optics and Mietronics, enabling numerous optical device applications~\cite{jahani_alldielectric_2016,KUZNETSOV_Science_optically_2016,LIU_2018_Opt.Express_Generalized,WON_2019_Nat.Photonics_Mietronic}. The merging of optically-induced magnetism of nonmagnetic materials with the electromagnetic duality symmetry would set free duality principles and their applications from magnetic materials, which can potentially bring the electromagnetic duality symmetry to the masses of the optics and photonics community~\cite{FERNANDEZ-CORBATON_2013_Phys.Rev.B_Role,RAHIMZADEGAN_2018_Phys.Rev.Applied_CoreShell,YANG_2020_ArXiv}.

In a previous study, we have blended optically-induced magnetism and the duality symmetry to demonstrate how the angular scattering patterns are invariant under duality transformations and how the backward scatterings and the associated reflections of infinite periodic structures can be eliminated when extra rotation symmetry is further exploited~\cite{YANG_2020_ArXiv}. In this work, in the same spirit, we shift our focus from direction dependent angular scattering patterns to integral total scattering properties (including extinction, scattering and absorption cross sections) of non-magnetic particle clusters, and reveal that all those cross sections are invariant under duality transformations. To be more specific, clusters made of self-dual particles exhibit polarization independent total scattering properties for any given incident direction; while for clusters consisting of non-self-dual particles, the total scattering properties are identical for any two scattering configurations that can be duality transformed into each other. Such total scattering invariance has nothing to do with specific geometric distributions of the clusters, being the clusters finite or infinite, random or periodic.  Such robust scattering features may not only play significant roles in fundamental scattering related studies involving electromagnetic dualtiy (such as topological and/or non-hermitian photonics), but also find applications in a wide range of optical devices that require optically stable functionalities.

\section{Duality transformations for dipolar particles}
\label{Duality}

In this work, we confine our studies to the plane wave scatting in a homogenous background (vacuum in this study) by clusters consisting of dipolar particles that support electric dipole (ED) moment \textbf{p} and magnetic dipole (ED) moment \textbf{m} (higher order multipolar moments are negligible). An electromagnetic duality transformation for the fields and moments can be expressed as~\cite{JACKSON_1998__Classical,FERNANDEZ-CORBATON_2013_Phys.Rev.B_Role,YANG_2020_ArXiv}:
%--------------------------------------------------------------
\begin{equation}\begin{array}{l}
\label{duality_transformation}
\left(\begin{array}{c}
\mathbf{E} \\
\mathbf{H}
\end{array}\right) \rightarrow\left(\begin{array}{c}
\mathbf{E}^{\prime} \\
\mathbf{H}^{\prime}
\end{array}\right)=T(\beta)\left(\begin{array}{l}
\mathbf{E} \\
\mathbf{H}
\end{array}\right), \\
\left(\begin{array}{c}
\mathbf{p} \\
\mathbf{m}
\end{array}\right) \rightarrow\left(\begin{array}{c}
\mathbf{p}^{\prime} \\
\mathbf{m}^{\prime}
\end{array}\right)=T(\beta)\left(\begin{array}{c}
\mathbf{p} \\
\mathbf{m}
\end{array}\right).
\end{array}\end{equation}
%-------------------------------------------------------------
Here $\mathbf{E}$ and $\mathbf{H}$ correspond to either incident or scattered waves; and $T(\beta)$ is the duality transformation matrix:
%--------------------------------------------------------------
\begin{equation}
\label{duality_rotation_matrix}
T(\beta)=\left[\begin{array}{ll}
\cos \beta, & -\sin \beta\\
\sin \beta, & ~~\cos \beta\end{array}\right],\end{equation}
%--------------------------------------------------------------
with a real transformation angle $\beta$. For the incident plane wave and the scattered waves in the far field, $\mathbf{E}$ and $\mathbf{H}$ are perpendicular
($\mathbf{E}\perp\mathbf{H}$) and then the duality transformation corresponds to a geometric rotation of angle $\beta$ along the wave propagation direction (unit Poynting vector $\hat{\mathbf{s}}$) $\check{R}_{\hat{\mathbf{s}}}(\beta)$~\cite{YANG_2020_ArXiv}. The same correspondence is also applicable to the electric moment \textbf{p} and the magnetic moment \textbf{m}, as along as they are perpendicular to each other.

When the dipolar responses of the scattering particles to the exciting fields  are fully isotropic, the fields and moments are connected by~\cite{JACKSON_1998__Classical,SERSIC_2011_Phys.Rev.B_Magnetoelectric,YANG_2020_ArXiv}:
%------------------------------------------------
\begin{equation}\left(\begin{array}{c}
\mathbf{p} \\
\mathbf{m}
\end{array}\right)=\left[\begin{array}{cc}
\alpha^{e}, & 0 \\
0, & \alpha^{m}
\end{array}\right]\left(\begin{array}{c}
\mathbf{E} \\
\mathbf{H}
\end{array}\right).\end{equation}
%-----------------------------------------------
Here $\alpha^{e}$ and $\alpha^{m}$ are scalars, corresponding respectively to the electric and magnetic polarizabilities. For isotropic dipolar particles characterized by Mie coefficients $a_1$ and $b_1$,  the polarizabilities can be expressed as: ${\alpha^{e}} = {{3i} \over {2{k^3}}}{a_1},~{\alpha^{m}} = {{3i} \over {2{k^3}}}{b_1}$~\cite{ Wheeler2006_PRB,Liu2012_ACSNANO,YANG_2020_ArXiv,Bohren1983_book}. To guarantee that after the duality transformation, the duality-paired scattering configuration still shows an isotropic response:
%==============================
\begin{equation}\left(\begin{array}{l}
\mathbf{p}^{\prime} \\
\mathbf{m}^{\prime}
\end{array}\right)=\left[\begin{array}{cc}
\alpha^{\prime e}, & 0 \\
0, & \alpha^{\prime m}
\end{array}\right]\left(\begin{array}{l}
\mathbf{E}^{\prime} \\
\mathbf{H}^{\prime}
\end{array}\right),
\end{equation}
%========================================
at least one of the following conditions should be satisfied~\cite{FERNANDEZ-CORBATON_2013_Phys.Rev.B_Role,YANG_2020_ArXiv}: (i) $\mathbf{p}=\mathbf{m}$ (self-dual dipolar particle $a_1=b_1$), $\alpha^{\prime e,m}= \alpha^{e,m}$ (or equivalently $a_1=b_1=b_1^\prime=a_1^\prime$), and $\beta$ is arbitrary; (ii) $\mathbf{p}\neq \mathbf{m}$ (non-self-dual dipolar particle $a_1\neq b_1$), $\alpha^{\prime e,m}= \alpha^{m,e}$ (or equivalently $a_1^\prime=b_1$, $b_1^\prime=a_1$), and $\beta=\pi/2$. Basically, for non-self-dual particles, the duality transformation would require the following Mie coefficients change: $(a_1, b_1)\rightarrow (b_1, a_1)$, meaning that the particle is transformed to its dual-partner; in contrast, the self-dual particle is kept as it is under the duality transformation, as its dual-partner is itself.

\section{Total scattering invariance under Duality transformations}
\label{Invariance}

Now we proceed to discuss how the total scattering properties are preserved under duality transformations. For an incident plane wave (fields are $\mathbf{E}_{i}$ and  $\mathbf{H}_{i}$; wave vector is $\mathbf{k}_{i}$) upon an arbitrary obstacle (could be single particles or clusters), all the cross sections of extinction, scattering and absorption can be extracted from the incident and scattered fields ($\mathbf{E}_{s}$ and  $\mathbf{H}_{s}$). To be specific, the extinction cross section can be expressed as~\cite{Bohren1983_book}:
%===================================
\begin{equation}
  \sigma_{\mathrm{ext}}=\frac{1}{I_i}\iint \mathrm{\mathbf{S}}_{\mathrm{ext}}\cdot \mathbf{\hat{e}}_r~d A, ~ \mathrm{\mathbf{S}}_{\mathrm{ext}}=\operatorname{Re}\left(\mathrm{\mathbf{E}}_{\mathrm{i}} \times \mathrm{\mathbf{H}}_{\mathrm{s}}^{*}+\mathrm{\mathbf{E}}_{\mathrm{s}} \times \mathrm{\mathbf{H}}_{\mathrm{i}}^{*}\right),
\end{equation}
%===================================
where the incident irradiance is $I_i=\mathbf{E}_i \mathbf{E}_i^{*}$; $\mathbf{\hat{e}}_r$ is the radial unit vector along which light is scattered; $\ast$ denotes the complex conjugation; $Re$ means the real part; and $A$ is a closed spherical surface enclosing the obstacle. According to the optical theorem, the extinction cross section can be further simplified as~\cite{Bohren1983_book}:
%====================
\begin{equation}
\sigma_{\mathrm{ext}}=\frac{4 \pi}{k^2} \operatorname{Re}\left[\mathbf{E}_{i}^{*} \cdot \mathbf{F} \left(\mathbf{k}=\mathbf{k}_{i}\right)\right],
\end{equation}
%====================
where $\mathbf{k}$ is the wave vector of the scattered wave and $k= |\mathbf{k}|$; $\mathbf{F}(\mathbf{k} = {\mathbf{k}_i})$ is vectorial scattering amplitude (normalized by the incident electric field amplitude)~\cite{Bohren1983_book}.  Since both the incident field and the forward scattered fields are transverse with $\mathbf{E} \perp \mathbf{H}$ and propagating along $\mathbf{k}_i$, a duality transformation simply corresponds to a rotation along $\mathbf{k}_i$
for both $\mathbf{F}(\mathbf{k} = {\mathbf{k}_i})$ and $\mathbf{E}_{i}$. Then the extinction cross section after the duality transformation would be
%%====================
\begin{equation}\begin{array}{c}
\sigma_{\mathrm{ext}}^{\prime}=\frac{4 \pi}{k^{2}} \operatorname{Re}\left[\mathbf{E}_{i}^{\prime*} \cdot \mathbf{F}^{\prime}\left(\mathbf{k}=\mathbf{k}_{i}\right)\right]= \\\\
\frac{4 \pi}{k^{2}} \operatorname{Re}\left[\left(\check{R}_{\mathbf{\hat{s}} \| \mathbf{k}_{i}}(\beta) \mathbf{E}_{i}\right)^{*} \cdot\left(\check{R}_{\mathbf{\hat{s}} \| \mathbf{k}_{i}}(\beta) \mathbf{F}\left(\mathbf{k}=\mathbf{k}_{i}\right)\right)\right],
\end{array}\end{equation}
%%==========================
which is obviously invariant: $\sigma_{\mathrm{ext}}^{\prime}=\sigma_{\mathrm{ext}}$, since the dot product of two vectors is a constant regardless of the orientation of the axis: $\mathbf{E}_{i}^{*} \cdot \mathbf{F}\left(\mathbf{k}=\mathbf{k}_{i}\right)=\left(\check{R}_{\mathbf{\hat{s}} \| \mathbf{k}_{i}}(\beta) \mathbf{E}_{i}\right)^{*} \cdot\left(\check{R}_{\mathbf{\hat{s}} \| \mathbf{k}_{i}}(\beta) \mathbf{F}\left(\mathbf{k}=\mathbf{k}_{i}\right)\right)$.

The scattering cross section can be expressed as~\cite{Bohren1983_book}:
%===================================
\begin{equation}
\sigma_{\mathrm{sca}}=\frac{1}{I_{i}} \iint \mathrm{\mathbf{S}}_{\mathrm{sca}} \cdot \hat{\mathbf{e}}_{r} d A,~~ \mathrm{\mathbf{S}}_{\mathrm{sca}}=\frac{1}{2} \operatorname{Re}\left(\mathbf{E}_{\mathrm{s}} \times \mathbf{H}_{\mathrm{s}}^{*}\right).
\end{equation}
%======================
In the far field, the scattered waves are transverse ($\mathbf{E}_{\mathrm{s}} \times \mathbf{H}_{\mathrm{s}}$) spherical waves with $\mathrm{\mathbf{S}}_{\mathrm{sca}} \| \hat{\mathbf{e}}_{r}$. In a similar fashion, it is easy to prove that the scattering cross section is also invariant $\sigma_{\mathrm{sca}}=\sigma_{\mathrm{sca}}^\prime$ under a duality transformation as:
%==============================
\begin{equation}
\mathbf{E}_{\mathrm{s}}^{\prime} \times \mathbf{H}_{\mathrm{s}}^{\prime *}=\left(\check{R}_{\mathrm{\hat{s}} \| \hat{\mathbf{e}}_{r}}(\beta) \mathbf{E}_{\mathrm{s}}\right) \times\left(\check{R}_{\mathrm{\hat{s}} \| \hat{\mathbf{e}}_{r}}(\beta) \mathbf{H}_{\mathrm{s}}\right)^{*}=\mathbf{E}_{\mathrm{s}} \times \mathbf{H}_{\mathrm{s}}^{*}.
\end{equation}
%==============================
This is simply due to the fact that the cross product of two perpendicular vectors is irrelevant to the in-plane (perpendicular to $\hat{\mathbf{e}}_{r}$) axis orientations. The invariance of both scattering and extinction cross sections would directly lead to the invariance of absorption cross section $\sigma_{\mathrm{abs}}$, as ${\sigma _{{\rm{abs   }}}} = {\sigma _{\rm{ext}}} - {\sigma _{\rm{sca}}}$ according to the optical theorem~\cite{Bohren1983_book}.

\section{Total scattering invariance for finite particle clusters}
\label{Finite}

\subsection{Self-dual particle clusters}

After the theoretical analysis above, we now turn to specific demonstrations with realist nonmagnetic particles.  For arbitrary clusters consisting of self-dual dipolar particles, it is shown in Section~\ref{Duality} that the particles are kept as they are under the duality transformation. The transformation of the incident and the scattered fields correspond to a geometric rotation around the propagating direction of an arbitrary angle $\beta$. This directly guarantees that for any fixed incident direction, all sorts of cross sections of the self-dual particle cluster would be independent of the polarization direction.

\begin{figure}
%\begin{center}
\includegraphics[width=8.2cm]{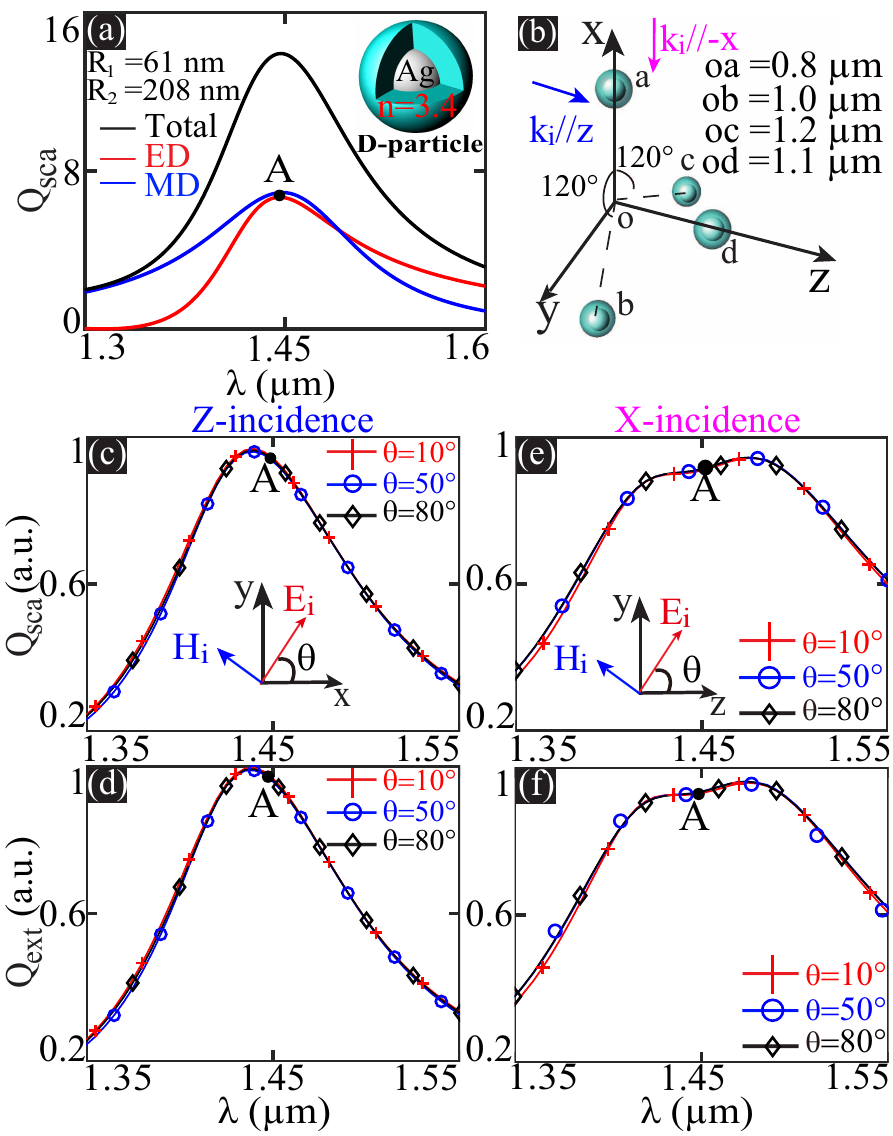}
%\end{center}
\caption{(a) Scattering eﬃciency spectra (both total scattering and partial dipolar contributions from ED and MD) for the Ag core-dielectric (In=3.4) shell spherical particle ( $R_1$ = 61 nm and $R_2$ = 208 nm).  The resonant position is marked at $\lambda_\mathbf{A} = 1449$ nm. (b) A randomly chosen particle cluster consisting of four such particles, with three particles on the \textbf{x-y} plane (one on the \textbf{x} axis and the other two with position vectors making the same angle of $120^\circ$ with respect to the \textbf{x} axis) and the forth particle is placed at the \textbf{z} axis. The distances between the particles and the coordinate origin  are $oa =800$ nm, $ob =1000$ nm, $oc = 1200$ nm and $od= 1100$ nm. (c) and (d): The scattering and extinction efficiency for three polarization directions ($\theta = 10^\circ, ~50^\circ, ~80^\circ $ with respect to the \textbf{x} axis) when the incident plane wave is propagating along \textbf{z} axis. (e) and (f): The efficiency for three polarization directions ($\theta = 10^\circ, ~50^\circ, ~80^\circ $ with respect to the \textbf{z} axis) when the incident  wave is propagating along \textbf{-x} axis.}
\label{fig1}
\end{figure}

To verify this polarization independent feature, we utilize the self-dual metal (Ag)-dielectric (refractive index In=3.4) core-shell spherical particle that has been widely studied in previous studies~\cite{Wheeler2006_PRB,Liu2012_ACSNANO,YANG_2020_ArXiv,Bohren1983_book}. The optical constants of silver are adopted from the experimental data in Ref.~\cite{Johnson1972_PRB}. The scattering efficiency (scattering cross section divided by the geometric section) spectra, including both total scattering and those contributed by ED ($a_1$) and MD ($b_1$), are shown in Fig. \ref{fig1}(a). The particle is of inner radius $R_1 = 61$~nm and outer radius $R_2 = 208$~nm, as shown in the inset of Fig. \ref{fig1}(a). It is clear that the particle is self-dual ( $a_1 \approx b_1$) at the resonant wavelength ${\lambda_\mathbf{A}} = 1449$~nm, and thus termed as $\mathbf{D}$-\textbf{Particle} (here \textbf{D} means dual) throughout this work.

As the total scattering invariance is irrelevant to specific geometric distributions of the particles, we randomly choose a cluster consisting of four \textbf{D}-\textbf{particles} as shown in Fig.~\ref{fig1}(b), and the position parameters are detailed in the caption of Fig.~\ref{fig1}. Two scenarios of different incident directions ($\mathbf{k}_i \| \mathbf{z}$ and $\mathbf{k}_i \| \mathbf{-x}$) are investigated, and for each case the results (obtained with commercial software COMSOL MULTIPHYSICS: www.comsol.com) for three polarization directions are shown [Figs.~\ref{fig1}(c)-(f)]. In Fig.~\ref{fig1} and throughout this work, we show only the results of extinction and scattering invariance, which directly guarantee the invariance of absorption according to the optical theorem. It is worth mentioning that the cross section invariance is observed for not just the designed spectral position ($\lambda_\mathbf{A}$ = 1449 nm), but rather across a broadband spectral regime, where the \textbf{D}-\textbf{particle} can be also viewed approximately as self-dual [see Fig.~\ref{fig1}(a)].  It is worth mentioning that the polarization-independent scattering features we have shown here totally originate from the self-duality of each consisting particle, and thus have nothing to do with specific geometric distributions or incident directions. The same polarization-independent scattering and absorption properties have also been achieved with obstacles that exhibit composite mirror and rotation symmetry $C_{nv}$ ($n \geq 3$)~\cite{Hopkins2013_nanoscale}, which nevertheless imposes rather stringent restrictions on the geometric distributions and the incident directions.

\subsection{Non-self-dual particle clusters}

As has been discussed in Section~\ref{Duality}, compared to its self-dual counterpart, the non-self-dual particle cluster is strikingly different in the sense that the duality transformation changes not only the fields but also the cluster itself~\cite{YANG_2020_ArXiv}. The scattering configuration before and after the transformation constitute a dual-pair, and if both configurations show pure isotropic response, then only the duality transformation of $\beta=\pi/2$ is allowed~\cite{YANG_2020_ArXiv}. This specific duality transformation would: (i) change the incident wave to its orthogonal polarization state with the incident direction fixed; (ii) transform the consisting particles according to  $\alpha^{\prime e,m}= \alpha^{m,e}$, and for dipolar particles it is equivalent to $a_1^\prime=b_1$, $b_1^\prime=a_1$, meaning a non-self-dual particle is transformed to its dual-partner.

%======================================================
\begin{figure}[t]
\begin{center}
\includegraphics[width=8.8cm]{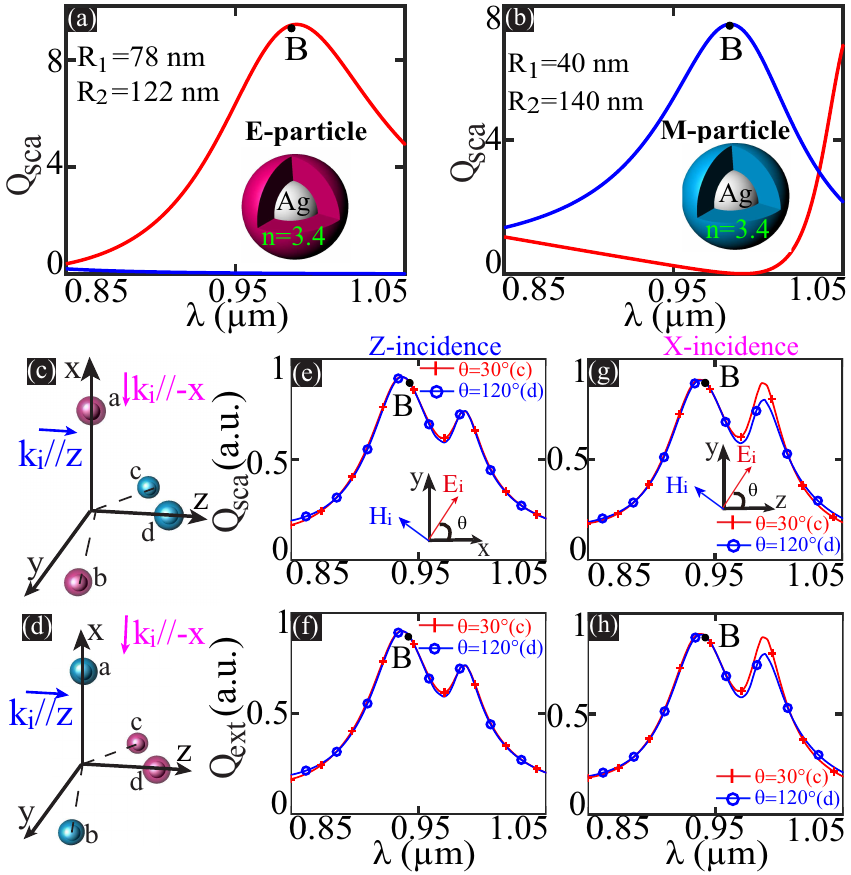}
\end{center}
\caption{Scattering eﬃciency spectra of (a) an $\mathbf{E}$-\textbf{Particle} with $R_1 = 78$ nm and $R_2 = 122$ nm, and (b) a $\mathbf{M}$-\textbf{Particle} with $R_1 =40$ nm and $R_2 =140$ nm. The two particles support respectively pure ED and MD moments at the marked resonant point $\lambda_\mathbf{B} =984$ nm.  (c) and (d): The dual-paired conﬁgurations with a duality transformation of $\beta = \pi/2$ (swapping the positions of $\mathbf{E}$-\textbf{Particles} and $\mathbf{M}$-\textbf{Particles}) with two different incident directions ($\mathbf{k}_i\| \mathbf{z}$ and $\mathbf{k}_i\| \mathbf{-x}$). The particle position parameters  are the same as those in Fig. \ref{fig1}(b). (e) and (f): The scattering and extinction efficiency for the incidence along \textbf{z} with polarization angles chosen as $\theta = 30^\circ$ in (c) and $120^\circ$ in (d). (g) and (h): The scattering and extinction efficiency for another incidence along \textbf{-x}.}
\label{fig2}
\end{figure}
%======================================================

%==========================================
\begin{figure}[t]
\includegraphics[width=8.8cm]{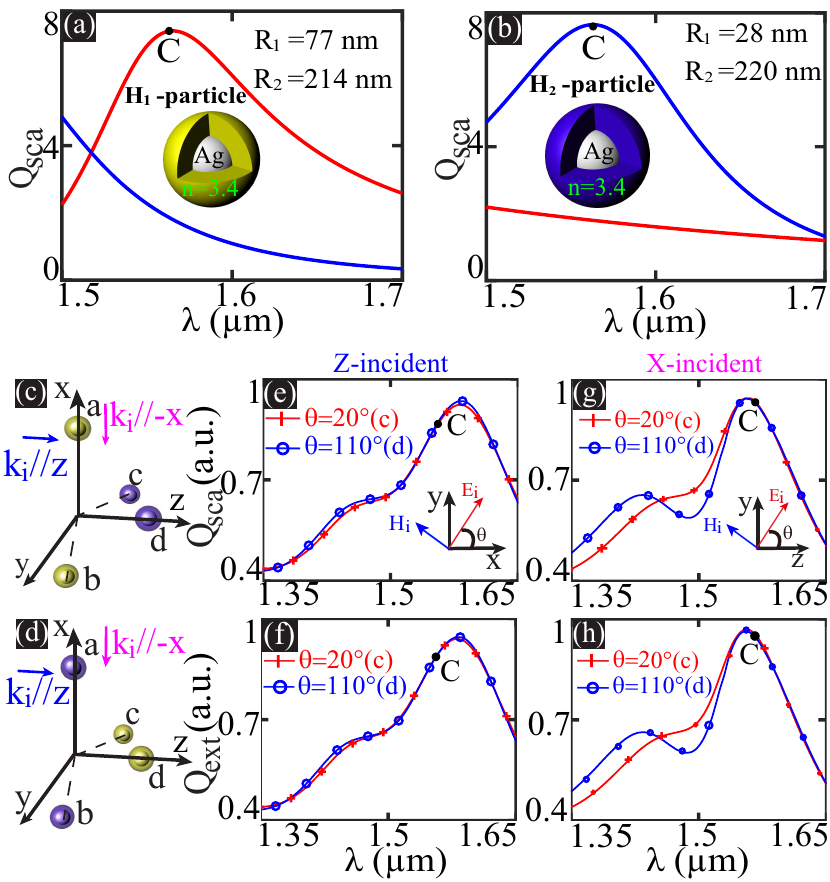}
\caption{Scattering eﬃciency spectra of (a) an $\mathbf{H_1}$-\textbf{Particle} with $R_1 = 77$ nm and $R_2 = 214$ nm, and (b) a $\mathbf{H_2}$-\textbf{Particle} with with $R_1 = 28$ nm and $R_2 = 220$ nm. The two particles support both ED and MD moments at the marked resonant point $\lambda_\mathbf{C} =1560$ nm.  (c) and (d): The dual-paired conﬁgurations with a duality transformation of $\alpha = \pi/2$ (swapping the positions of $\mathbf{H_1}$-\textbf{Particles} and $\mathbf{H_2}$-\textbf{Particles}) with two different incident directions ($\mathbf{k}_i\| \mathbf{z}$ and $\mathbf{k}_i\| \mathbf{-x}$). The position parameters  are the same as those in Fig. \ref{fig1}(b). (e) and (f): The scattering and extinction efficiency for the incidence along \textbf{z} with polarization angles chosen as $\theta = 20^\circ$ in (c) and $110^\circ$ in (d). (g) and (h) The scattering and extinction efficiency for another incidence along \textbf{-x}.}
\label{fig3}
\end{figure}
%==============================================

The simplest dual-partners would consist of an ED particle ($\mathbf{E}$-\textbf{Particle}: $a_1\neq 0$, $b_1=0$) and its dual-partner a MD particle ($\mathbf{M}$-\textbf{Particle}: $a_1^\prime=0$, $b_1^\prime=a_1$).  We show the scattering eﬃciencies of such dual-partners in Figs. \ref{fig2}(a) and (b): an $\mathbf{E}$-\textbf{Particle} of $R_1 =78$ nm and $R_2 = 122$ nm and a $\mathbf{M}$-\textbf{Particle}  with $R_1 = 40$ nm and $R_2 = 140$ nm. Both particles are made of Ag core and dielectric-shell (In=3.4) and support either a pure ED or MD at the common marked resonant wavelength $\lambda_\mathbf{B} = 984$ nm. Two dual-paired scattering configurations consisting of  $\mathbf{E}$, $\mathbf{M}$-\textbf{Particles} are shown in Figs. \ref{fig2}(c) and (d), where the $\mathbf{E} (\mathbf{M})$-\textbf{Particles} in Fig. \ref{fig2}(c) are replaced by their dual-partnered $\mathbf{M} (\mathbf{E})$-\textbf{Particles}. The overall cluster configuration and the particle position parameters are the same as those in Fig. \ref{fig1}(b). The extinction and scattering efficiencies for two different incident directions are summarized in Figs. \ref{fig2}(e)-(h), and for each direction we have arbitrary chosen two orthogonal polarizations ($\theta = 30^\circ$ and $120^\circ$) as required by $\beta=\pi/2$. Similar to what is shown in Figs. \ref{fig1}(c)-(f), the total scattering invariance can be observed at not only the designed spectral position, but across a relatively broad spectral regime, where the two particles can be treated approximately as dual-partners.
In Figs. \ref{fig2}(g)-(h), the scattering invariance breaking regime is clearly visible (around $\lambda=1045$~nm), where the two particles are not dual-partners anymore.

The more general dual-partners would be hybrid particles that support both electric and magnetic moments ($a_{1}^\prime=b_{1}\neq 0$, $b_{1}^\prime=a_{1}\neq 0$). We show the scattering eﬃciencies of such dual-partners in Figs. \ref{fig3}(a) and (b): a $\mathbf{H_1}$-\textbf{Particle} of $R_1 = 77$ nm and $R_2 = 214$ nm and a $\mathbf{H_2}$-\textbf{Particle}  with $R_1 = 28$ nm and $R_2 = 220$ nm. Both particles are of Ag-core and dielectric-shell (In=3.4) and support both ED and MD moments at the resonant wavelength $\lambda_\mathbf{C} = 1560$ nm. Two dual-paired scattering configurations consisting of  $\mathbf{H_1}$, $\mathbf{H_2}$-\textbf{Particles} are shown in Figs. \ref{fig3}(c) and (d), where the $\mathbf{H_1} (\mathbf{H_2})$-\textbf{Particles} in Fig. \ref{fig3}(c) are replaced by their dual-partner $\mathbf{H_2} (\mathbf{H_1})$-\textbf{Particles}. The cluster configuration and particle position parameters are the same as those in Fig. \ref{fig1}(b). The extinction and scattering efficiencies for two different incident directions are summarized in Figs. \ref{fig3}(e)-(h), and for each direction we have arbitrary chosen two orthogonal polarizations ($\theta = 20^\circ$ and $110^\circ$). Similar to what is shown in Figs. \ref{fig1}(c)-(f) and Figs. \ref{fig2}(e)-(h), for the incidence along \textbf{z} axis, the total scattering invariance can be observed across a relatively broad spectral regime. While for the incidence along \textbf{-x} axis the scattering invariance is generally broken. For both scenarios, the total scattering invariance can be observed at the designed wavelength $\lambda_\mathbf{C} = 1560$ nm,  where the two particles are strictly dual partners.

%==============================================
\begin{figure}
\includegraphics[width=8.8cm]{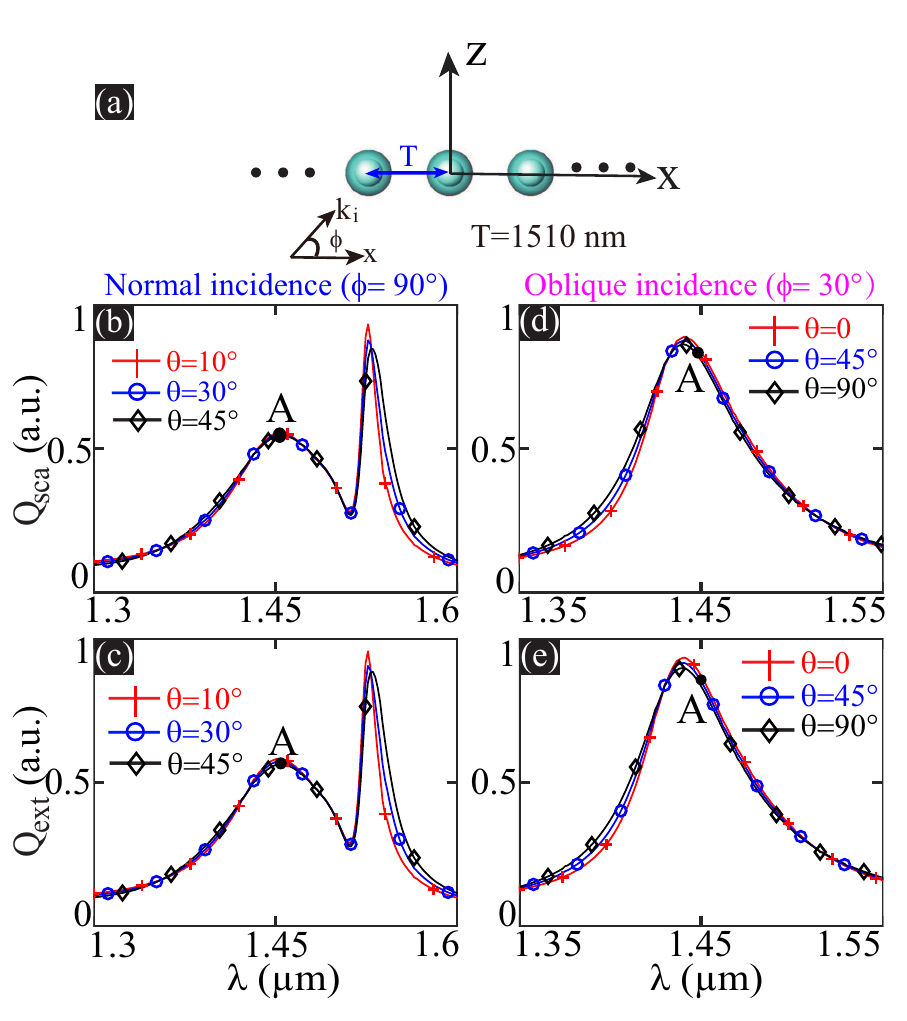}
\caption{(a) A 1D lattice of $\mathbf{D}$-\textbf{Particles} with period $T=1510$~nm. Extinction and scattering efficiency spectra are shown in (b) and (c) for a normal incidence ($\mathbf{k}_i\| \mathbf{z}$) with polarization angle $\theta = 10^\circ, ~30^\circ, ~45^\circ$, and in (d) and (e) for an oblique incidence ($\phi$ =30$^\circ$) with $\theta = 0, ~45^\circ, ~90^\circ$. The marked point A corresponds to the resonant wavelength ($\lambda_\mathbf{A} = 1449$~nm)  of the $\mathbf{D}$-\textbf{Particle}.}
\label{fig4}
\end{figure}
%============================================

\section{Total scattering invariance for infinite periodic lattices}
\label{Infinite}

\subsection{Self-dual particle lattices}
The principles we have revealed in Sections~\ref{Duality} and ~\ref{Invariance} are applicable to not only finite particle clusters, but also to clusters made of an infinite number of particles, such as the periodic particle lattices that play essential roles in various photonic applications. We start with a one-dimensional (1D) self-dual $\mathbf{D}$-\textbf{Particle} lattice shown schematically in Fig.~\ref{fig4}(a) with period $T=1510$~nm.  A similar lattice has been studied in a previous study, where it is shown that under normal incidence (incident direction perpendicular to the lattice axis), the extinction efficiency spectra exhibit typical polarization independent Fano resonances~\cite{Liu2012_PRB}. It was not understood then but now well clarified in this work that such polarization independence originates from the self-duality of each consisting particle in the lattice. In Figs.~\ref{fig4}(b)-(c) we reproduce the  spectra (for three arbitrarily chose polarization angles $\theta = 10^\circ, ~30^\circ, ~45^\circ$) of the normal incidence $\mathbf{k}_i\| \mathbf{z}$, and as expected the polarization independent Fano resonances are observed in terms of both extinction and scattering spectra. Here $\theta$ is the angle made by $\mathbf{E}_i$ with respect to the $\mathbf{x}-\mathbf{z}$ plane. Similar to what is shown in Figs.~\ref{fig1}(c)-(f), such independence is manifest not only in the designed wavelength $\lambda_\mathbf{A} = 1449$ nm, but rather across a broad spectral regime. We make a step further to show the spectra (for three arbitrarily chose polarization angles $\theta = 0, ~45^\circ, ~90^\circ$) in Figs.~\ref{fig4}(d)-(e) of an oblique incidence ($\mathbf{k}_i$ makes an angle of $\phi$ =30$^\circ$ with respect to the $\mathbf{x}$ axis). Though the Fano resonances disappear when the incident light is tilted~\cite{Markel2005_JPBMO}, the polarization independence feature is well protected by the electromagnetic duality symmetry.

\subsection{Non-self-dual particle lattices}

\begin{figure}[t]
\includegraphics[width=8.8cm]{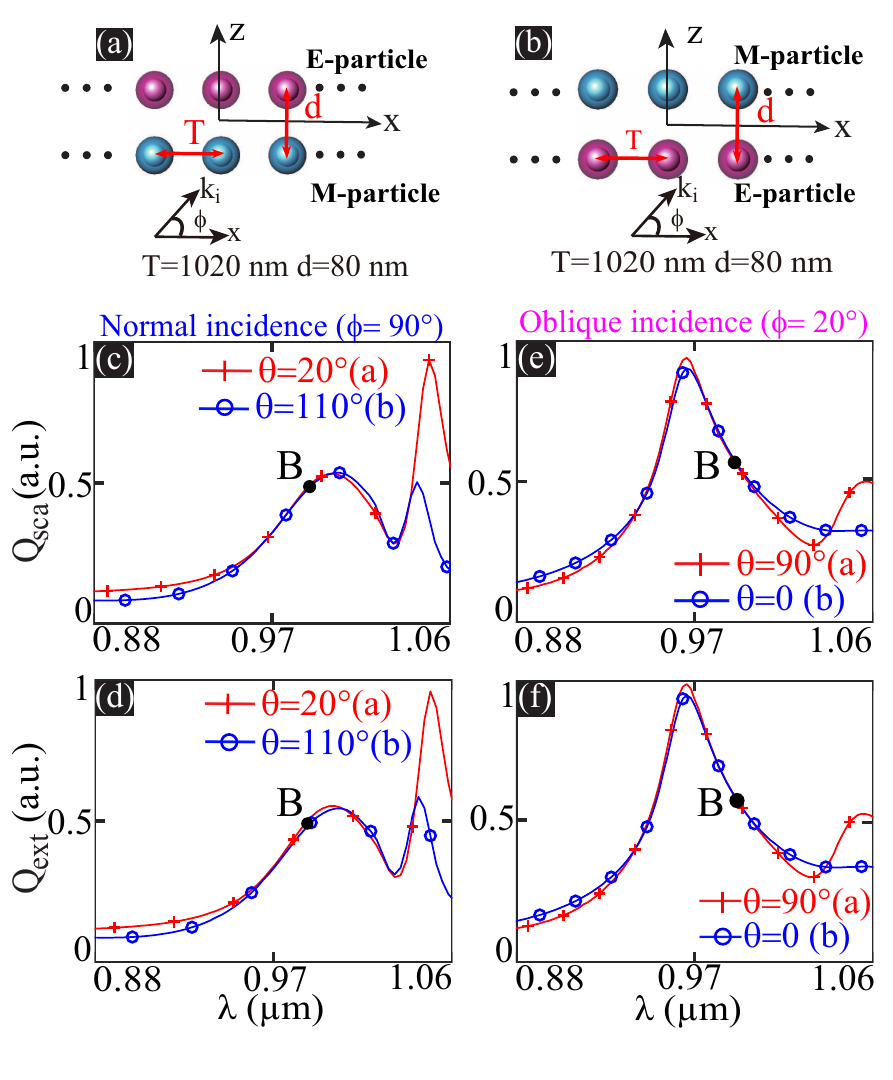}
\caption{(a) and (b): Two dual-paired 1D lattices consist of $\mathbf{E}$, $\mathbf{M}$-\textbf{Particles} of $T=1020$~nm and $d =80$ nm. Extinction and scattering efficiency spectra are shown: in (c) and (d) for a normal incidence ($\mathbf{k}_i\| \mathbf{z}$), with polarization angle $\theta = 20^\circ$ and $110^\circ$ for the configurations in (a) and (b), respectively; in (e) and (f) for an oblique incidence ($\phi$ =20$^\circ$), with polarization angle $\theta = 90^\circ$ and $0$ for (a) and (b), respectively.  The marked point \textbf{B} corresponds to the common resonant wavelength ($\lambda_\mathbf{B} = 984$~nm)  of the $\mathbf{E}$, $\mathbf{M}$-\textbf{Particles}.}
\label{fig5}
\end{figure}

\begin{figure}[t]
\includegraphics[width=8.8cm]{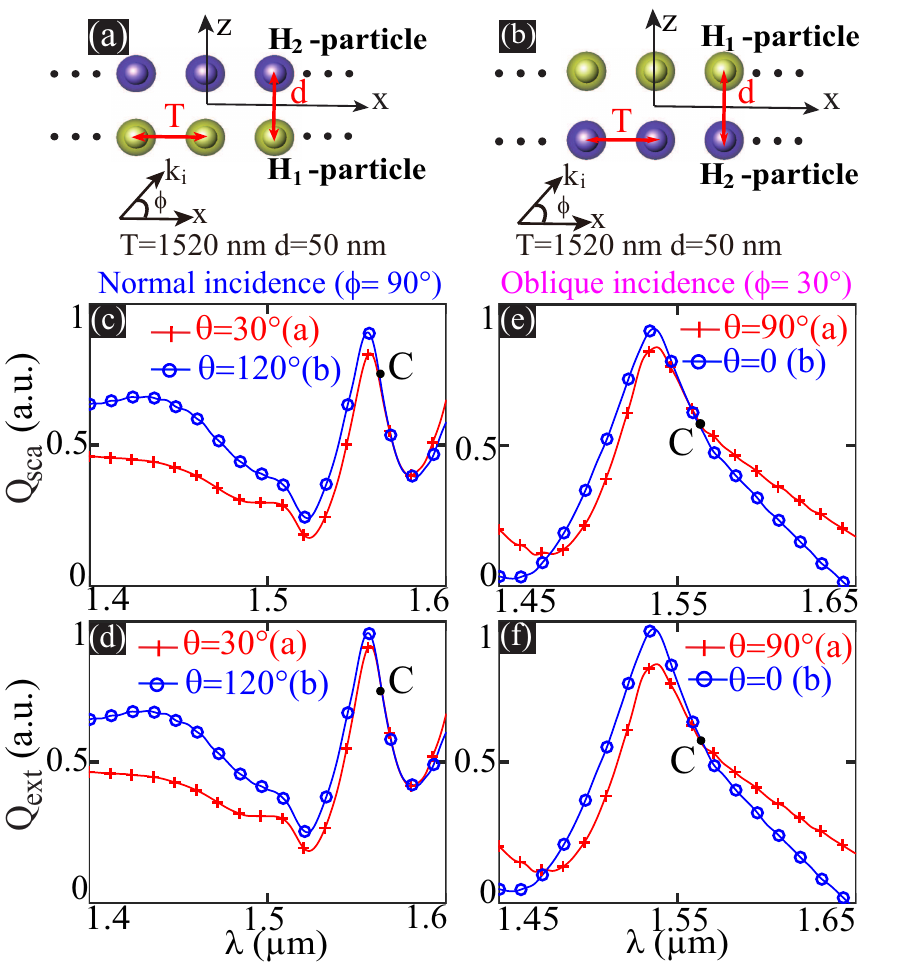}
\caption{(a) and (b): Two dual-paired 1D lattices consist of $\mathbf{H_1}$, $\mathbf{H_2}$-\textbf{Particles} of $T=1520$~nm and $d =50$ nm. Extinction and scattering efficiency spectra are shown: in (c) and (d) for a normal incidence ($\mathbf{k}_i\| \mathbf{z}$), with polarization angle $\theta = 30^\circ$ and $120^\circ$ for the configurations in (a) and (b), respectively;  in (e) and (f) for an oblique incidence ($\phi$ =30$^\circ$), with polarization angle $\theta = 90^\circ$ and $0$ for (a) and (b), respectively. The marked point \textbf{C} corresponds to the common resonant wavelength ($\lambda_\mathbf{C} = 1560$~nm)  of the $\mathbf{H_1}$, $\mathbf{H_2}$-\textbf{Particles}.}
\label{fig6}
\end{figure}

As a next step, we extend the study to 1D lattices of non-self-dual particles. Two 1D lattices (consisting of $\mathbf{E}$, $\mathbf{M}$-\textbf{Particles}) that are dual-partners to each other are shown in Figs.~\ref{fig5}(a) and (b): the period $T=1020$~nm; the inter-particle distance within the unit-cell is $d =80$ nm. Extinction and scattering efficiency spectra are shown: in Figs.~\ref{fig5}(c) and (d) for a normal incidence ($\mathbf{k}_i\| \mathbf{z}$), with polarization angle $\theta = 20^\circ$ and $110^\circ$ for the configurations in Figs.~\ref{fig5}(a) and (b), respectively;  in Figs.~\ref{fig5}(e) and (f) for a oblique incidence ($\phi$ =20$^\circ$), with polarization angle $\theta = 90^\circ$ and $0$ in Figs.~\ref{fig5}(a) and (b), respectively.  The polarization directions are chosen arbitrarily but are kept orthogonal to make sure the scattering configurations in  Figs.~\ref{fig5}(a) and (b) constitute a dual-pair. In a similar way, the extinction and scattering spectra for another set of dual-paired 1D lattices (consisting of $\mathbf{H_1}$, $\mathbf{H_2}$-\textbf{Particles}; $T=1520$~nm; $d =50$~nm) are summarized in Figs.~\ref{fig6}, including results of both normal and oblique incidences with two orthogonal polarizations for the dual-paired scattering configurations.  Figures.~\ref{fig5} and \ref{fig6} can verify the invariance of total scattering properties under duality transformations, as shown by the spectra overlap at the marked point $B$  ($\lambda_\mathbf{B} = 984$~nm) in  Fig.~\ref{fig5} and $C$ ($\lambda_\mathbf{C}= 1560$~nm) in Fig.~\ref{fig6}.

\section{Conclusion}

In conclusion, we have revisited, from the composite perspectives of optically-induced magnetism and electromagnetic duality, the classical problem of plane wave scatterings by nonmagnetic particle clusters of random distributions. It is revealed that generally the total scattering properties, including cross sections of extinction, scattering and absorption, are all invariant under duality transformations. For arbitrary clusters consisting of self-dual particles, the total scatterings are polarization independent for any fixed incident direction, irrespective of how the particles are distributed; for clusters made of non-self-dual particles, the total scattering properties are identical for two dual-paired scattering configurations. Such scattering invariance are manifest for both finite clusters and infinite lattices, which may not only spawn practical applications in robust optical devices, but also render new insights for fundamental scattering-related studies, such as topological, non-hermitian and quantum photonics.

In this study, we have investigated only particles that support lowest order electric and magnetic dipoles, and similar studies can certainly be extended to higher order multipoles when self-dual Mie particles are characterized by $a_n=b_n~(n\geq2)$. For the demonstrations with infinite periodic lattices, we have only shown the results of 1D case, and for higher dimensional case the total scattering invariance are surely preserved, as the arguments presented in Sections~\ref{Duality} and \ref{Invariance} are valid regardless of the dimension of the particle distribution. Throughout this work, though we have employed only the ideal core-shell spherical particles that can be analytically calculated, the principles revealed are generic and can be applied for any structures that are more accessible experimentally.

%\begin{acknowledgement}
We acknowledge the financial support from National Natural Science Foundation of China (Grant No. 11874026 and 11874426).
%\end{acknowledgement}

%\bibliographystyle{osajnl}
%\bibliography{References_scattering3}% Produces the bibliography via BibTeX.

\end{document}